\documentclass{aa}

\usepackage{graphicx}
\usepackage{txfonts}
\usepackage{color}
\usepackage{xspace}
\usepackage[multiple]{footmisc}

\usepackage{url}

\begin{document} 

   \title{GRBAlpha: the smallest astrophysical space observatory}
   \subtitle{Part 1 -- Detector design, system description and satellite operations}
    
   \author{Andr\'as P\'al\inst{1}
           \and
           Masanori Ohno\inst{2}
           \and
           L\'aszl\'o M\'esz\'aros\inst{1}
           \and
           Norbert Werner\inst{3}
           \and
           Jakub \v{R}\'{\i}pa\inst{3}
           \and
           Bal\'azs Cs\'ak\inst{1}
           \and
           Marianna Daf\v{c}\'{\i}kov\'a\inst{3}
           \and
           Marcel Frajt\inst{4}
           \and
           Yasushi Fukazawa\inst{2}
           \and
           Peter Han\'ak\inst{5}
           \and
           J\'an Hudec\inst{4}
           \and
           Nikola Hus\'arikov\'a\inst{3}
           \and
           Jakub Kapu\v{s}\inst{4}
           \and
           Miroslav Kasal\inst{6}
           \and
           Martin Kol\'a\v{r}\inst{3}
           \and
           Martin Koleda\inst{7}
           \and
           Robert Laszlo\inst{7}
           \and
           Pavol Lipovsk\'{y}\inst{5}
           \and
           Tsunefumi Mizuno\inst{2}
           \and
           Filip Münz\inst{3}
           \and
           Kazuhiro Nakazawa\inst{8}
           \and
           Maksim Rezenov\inst{4}
           \and
           Miroslav \v{S}melko\inst{9}
           \and
           Hiromitsu Takahashi\inst{2}
           \and
           Martin Topinka\inst{10}
           \and
           Tom\'a\v{s} Urbanec\inst{6}
           \and
           Jean-Paul Breuer\inst{3}
           \and
           Tam\'as Boz\'oki\inst{11}
           \and
           Gergely D\'alya\inst{12}
           \and
           Teruaki Enoto\inst{13}         
           \and
           Zsolt Frei\inst{14}
           \and
           Gergely Friss\inst{14}
           \and
           G\'abor Galg\'oczi\inst{14,15}
           \and
           Filip Hroch\inst{1}
           \and
           Yuto Ichinohe\inst{16}
           \and
           Korn\'el Kap\'as\inst{17,18,15}
           \and
           L\'aszl\'o L. Kiss\inst{1}
           \and
           Hiroto Matake\inst{2}
           \and
           Hirokazu Odaka\inst{19}
           \and
           Helen Poon\inst{2}
           \and
           Ale\v{s} Povala\v{c}\inst{6}
           \and
           J\'anos Tak\'atsy\inst{14,15}
           \and
           Kento Torigoe\inst{2}
           \and
           Nagomi Uchida\inst{20}
           \and
           Yuusuke Uchida\inst{21}
          }
   \institute{Konkoly Observatory, Research Centre for Astronomy and Earth Sciences, Konkoly-Thege M. \'ut 15-17, H-1121, Budapest, Hungary; \email{apal@szofi.net} 
   \and
   Hiroshima University, School of Science, Higashi-Hiroshima, Japan 
   \and
   Department of Theoretical Physics and Astrophysics, Faculty of Science, Masaryk University,  Brno, Czech Republic 
   \and
   Spacemanic Ltd, Bratislava, Slovakia 
   \and
   Faculty of Aeronautics, Technical University of Ko\v{s}ice, Slovakia 
   \and
   Department of Radio Electronics, Faculty of Electrical Engineering and Communication, Brno University of Technology, Brno, Czech Republic 
   \and
   Needronix Ltd, Bratislava, Slovakia 
   \and
   Department of Physics, Nagoya University, Nagoya, Aichi, Japan 
   \and
   EDIS vvd., Košice,  Slovakia
   \and
   INAF - Istituto di Astrofisica Spaziale e Fisica Cosmica, Milano, Italy 
   \and
   Institute of Earth Physics and Space Science (EPSS), Sopron, Hungary 
   \and
   Department of Physics and Astronomy, Universiteit Gent, Ghent, Belgium 
   \and
   School of Science, Kyoto University, Kyoto, Japan 
   \and
   E\"otv\"os Lor\'and University, Budapest, Hungary 
   \and
   Wigner Research Centre for Physics, Budapest, Hungary 
   \and
   Department of Physics, Rikkyo University, Tokyo, Japan 
   \and
    Department of Theoretical Physics, Institute of Physics, Budapest University of Technology and Economics, Budapest, Hungary 
    \and
    MTA-BME Quantum Dynamics and Correlations Research Group, Budapest University of Technology and Economics, Budapest, Hungary 
   \and
    Department of Earth and Space Science, Osaka University, Toyonaka, Osaka, Japan 
    \and
    Institute of Space and Astronautical Science, Japan Aerospace Exploration Agency, Japan 
    \and
    Tokyo University of Science, Noda, Chiba, Japan 
}

\def\grbalpha{\textit{GRBAlpha}\xspace}
\def\iic{\ensuremath{\rm I^2C}\xspace}
\def\refmark#1{#1}

\offprints{A. P\'al, \email{apal@szofi.net}}
\date{Received date / Accepted date}

 
\abstract
{}
{Since launched on 2021 March 22, the 1U-sized CubeSat \grbalpha operates and collects scientific data on high-energy transients, making it the smallest astrophysical space observatory to date. \grbalpha is an in-obit demonstration of a gamma-ray burst (GRB) detector concept suitably small to fit into a standard 1U volume. As it was demonstrated in a companion paper, \grbalpha adds significant value to the scientific community with accurate characterization of bright GRBs, including the recent outstanding event of GRB\,221009A.}
{The GRB detector is a $75\times75\times5$ mm CsI(Tl) scintillator frapped in a reflective foil (ESR) read out by an array of SiPM detectors, multi-pixel photon counters by Hamamatsu, driven by two separate, redundant units. To further protect the scintillator block from sunlight and protect the SiPM detectors from particle radiation, we apply a multi-layer structure of Tedlar wrapping, anodized aluminium casing and a lead-alloy shielding on one edge of the assembly. The setup allows observations of gamma radiation within the energy range of $70-890\,{\rm keV}$ with an energy resolution of $\sim30$\%.}
{Here, we summarize the system design of the \grbalpha mission, including the electronics and software components of the detector, some aspects of the platform as well as the current way of semi-autonomous operations. In addition, details are given about the raw data products and telemetry in order to encourage the community for expansion of the receiver network for our initiatives with \grbalpha and related experiments.}
{}

\maketitle


\section{Introduction}

\grbalpha is an in-orbit demonstration mission of a gamma detector system suitably small to fit into an 1U CubeSat size, having an approximate dimensions of $10\times10\times11\,{\rm cm}$. In this experiment, we validate our concept of employing such small detector system for extracting astrophysical data related to gamma-ray bursts \citep[GRBs,][]{pal2020, ripa2022a}. One of the most recent findings of \grbalpha was the characterization of GRB\,221009A \citep{veres2022,lesage2022}, an exceptionally bright and long gamma-ray burst reported first by Fermi Gamma-ray Burst Monitor (GBM). We note here that this event was also detected by a series of other instruments, including
AGILE/GRID \citep{Piano2022},
AGILE/MCAL \citep{Ursi2022},
{BepiColombo}/MGNS \citep{Kozyrev2022},
{Insight}-HXMT \& {SATech-01}/GECAM-C  \citep[HEBS;][]{An2023}, 
{INTEGRAL}/SPI-ACS \citep{Gotz2022},
Konus-{WIND} \& {SRG}/ART-XC \citep{Frederiks2023}, 
MAXI \& NICER \citep{Williams2023}, 
{Solar Orbiter}/STIX \citep{Xiao2022},
{STPSat-6}/SIRI-2 \citep{Mitchell2022}, and
\textit{XMM-Newton} \citep{Tiengo2023}.
As a comparatively small detector, \grbalpha provided an unsaturated observation \citep{ripa2022b} and therefore allowed the scientific community to accurately obtain the peak flux of the event (\v{R}\'{\i}pa et al., A\&A, accepted).

In this paper, we present a description of the detector subsystem, the satellite platform, the operations scheme and the data downlink management. With its mission concept, involvement of students and implementation of on-board transponder features, the satellite gained the support of the radio amateur community and had an International Amateur Radio Union (IARU) coordination for downlink telemetry frequency in the UHF band\footnote{\url{http://www.amsatuk.me.uk/iaru/finished_detail.php?serialnum=745}}. Such a world-wide community can be extremely valuable for GRB astrophysics due to the low latency of data downlink for various types of orbits. In order to meet our commitments towards the amateur radio community, an extensive description is also included in this paper about the data format related to the telemetry structure and the process required to convert raw data streams into a scientifically relevant format. In order to further extend the available data types for downlink and upgrade the scientific on-board software in accordance, free code points are still available in the data stream to preserve backward compatibility and attain a forward compatibility in the ground segment components. 

The structure of the paper is as follows. In Sec.~\ref{sec:detector} we describe the detector structure used for monitoring GRBs -- including the mechanical configuration of the detector, the analog and digital components of the electronics, the on-board digital signal processing (DSP) scheme as well as the structure of the data streams provided by the DSP block as it is saved on-board and retrieved from the satellite. Sec.~\ref{sec:satelliteplatform} details the core components of the satellite platform while Sec.~\ref{sec:operations} details the currently implemented operations scheme, including data formats used for downlink. Results from the commissioning of this satellite are detailed in Sec.~\ref{sec:commissioning} while we give a brief summary and plans about the future on-board payload software upgrade in Sec.~\ref{sec:summaryandplans}. 

\begin{table*}[!ht]
\centering
\begin{tabular}{l|l|l|l}
Integer range & Bytes & Bit pattern & Overlong range \\
\hline
$0\dots63$ & 1 & 00xxxxxx & -- \\
$64\dots2^{12}-1$ & 2 & 010xxxxx 1xxxxxxx & $0\dots63$ \\
$2^{12}\dots2^{18}-1$ & 3 & 0110xxxx 1xxxxxxx 1xxxxxxx & $0\dots2^{12}-1$, $0\dots63$ \\
$2^{18}\dots2^{24}-1$ & 4 & 01110xxx 1xxxxxxx 1xxxxxxx 1xxxxxxx & $0\dots2^{18}-1$, $0\dots2^{12}-1$, $0\dots63$ \\
$2^{24}\dots2^{30}-1$ & 5 & 011110xx 1xxxxxxx 1xxxxxxx 1xxxxxxx 1xxxxxxx & $0\dots2^{24}-1$, \dots, $0\dots63$  \\
Unused code points & 1+ & 011111xx [\dots] & -- \\
\end{tabular}\vspace*{3mm}
\caption{The self-synchronizing variable-length coding as employed in the scientific data streams of \grbalpha on-board units and found also in the telemetry format. Up to 30-bit integers are supported at the present moment which also allows integers up to 24-bit length to be encoded in the overlong range. However, 2 bits of code point space are still available for arbitrary future extensions in order to ensure forward compatibility.}
\label{tab:variablelengthencoding}
\end{table*}

\begin{table*}[!ht]
\centering
\begin{tabular}{l|l}
Integer value or range & Interpretation \\[1pt]
\hline
2-byte overlong $3\dots31$ & consecutive zeros, number of zeros is the value (i.e. between $3$ and $31$)\\
2-byte overlong $32\dots39$ & spectrum, bin mode is the value minus 32 (i.e. between 0 and 7) \\
2-byte overlong $40$ & absolute timing, followed by 3 integers: seconds (upper 8 bits and lower 24 bits) and microseconds \\
2-byte overlong $41$ & relative timing, followed by a single integer (microseconds, w.r.t. the previous timestamp) \\
2-byte overlong $42$ & metadata \& housekeeping data: index, exposure time in $\mu$s, total count, cutoff value, temperatures \\
2-byte overlong $42\dots47$ & reserved for future housekeeping information and metadata of satellite platform components \\
2-byte overlong $48\dots63$ & reserved for future synchronization patterns \\
3-byte overlong $32\dots255$ & consecutive zeros, number of zeros is the value (i.e. between $32$ and $255$) \\
4-byte overlong values & unallocated code points, reserved for future use \\
5-byte overlong values & unallocated code points, reserved for future use
\end{tabular}\vspace*{3mm}
\caption{Data stream synchronization patterns and their respective interpretations for integers encoded in an overlong form. All of the overlong code space can also be used for self-synchronization purposes. A portion of the code space is used to encode a longer series of zeros (found in the high resolution calibration spectra taken in low-background regions), precise absolute timestamps, precise differential timestamps, housekeeping data, metadata associated with the data acquisition parameters and another types of synchronization patterns.}
\label{tab:syncpatterns}
\end{table*}

\begin{figure}
\centering
\includegraphics[width=85mm]{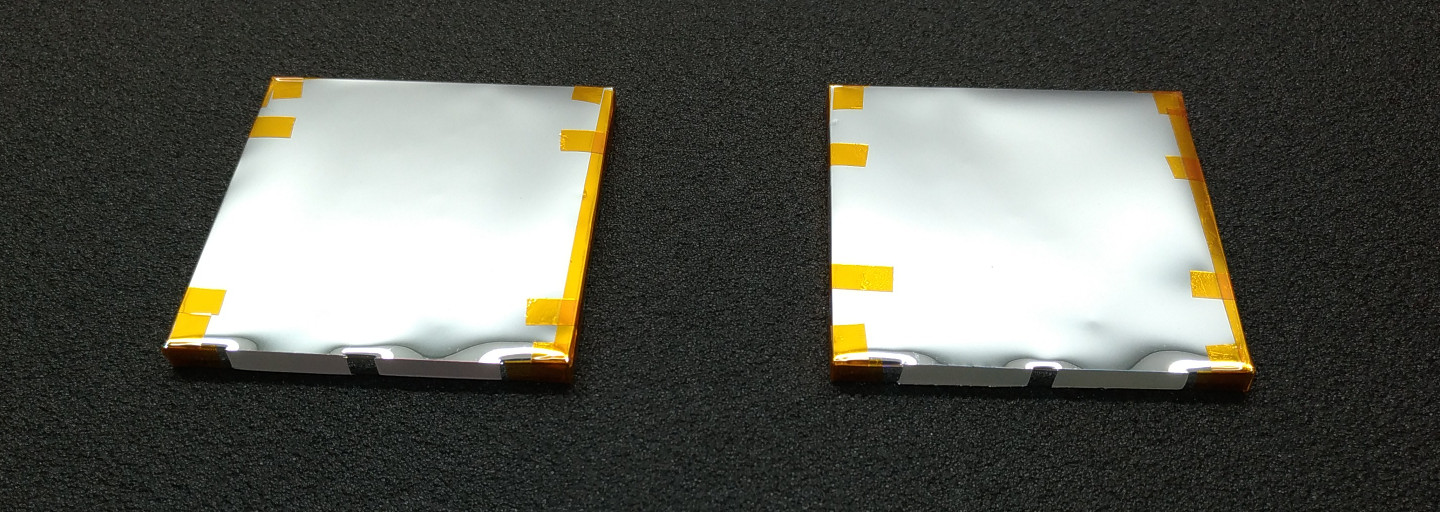}
\includegraphics[width=85mm]{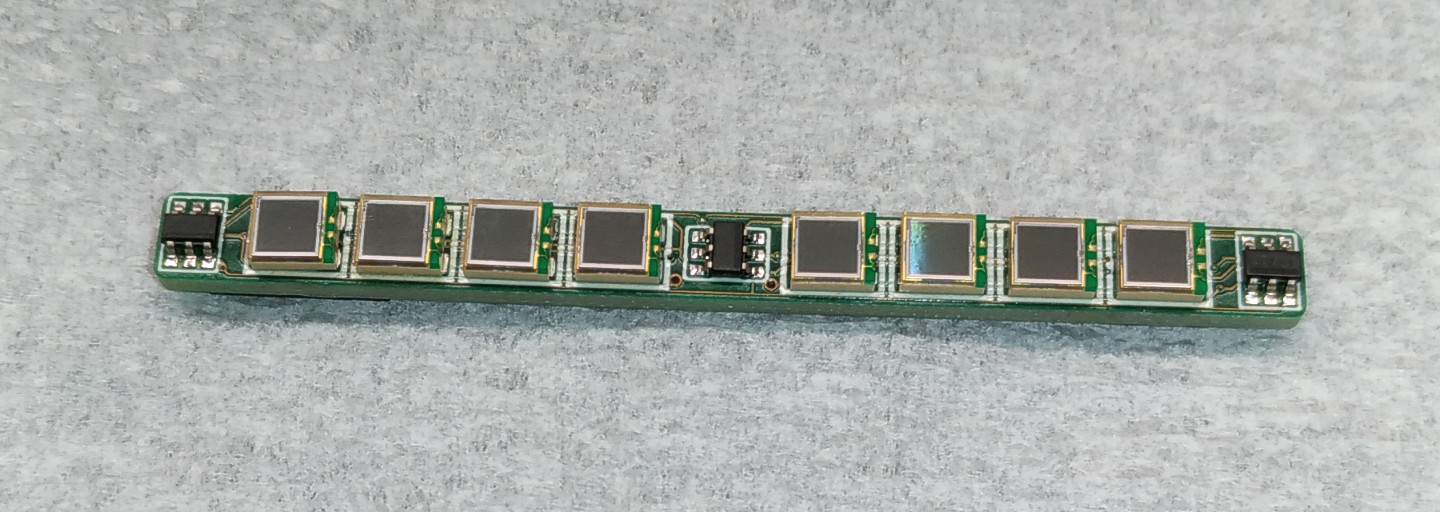}
\includegraphics[width=85mm]{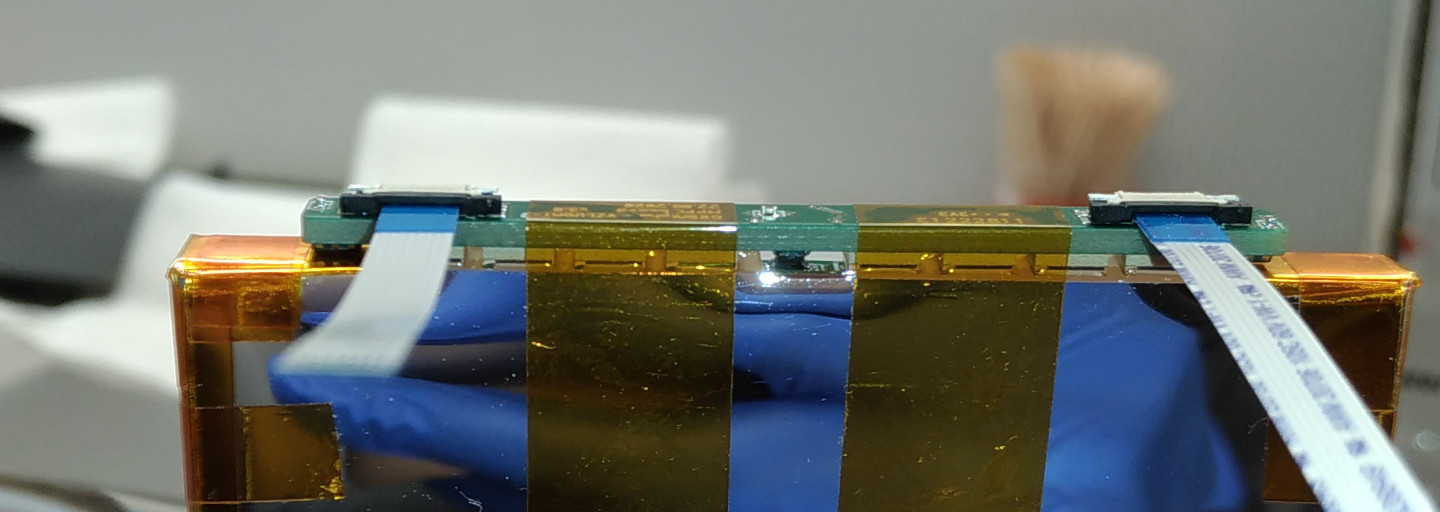}
\includegraphics[width=85mm]{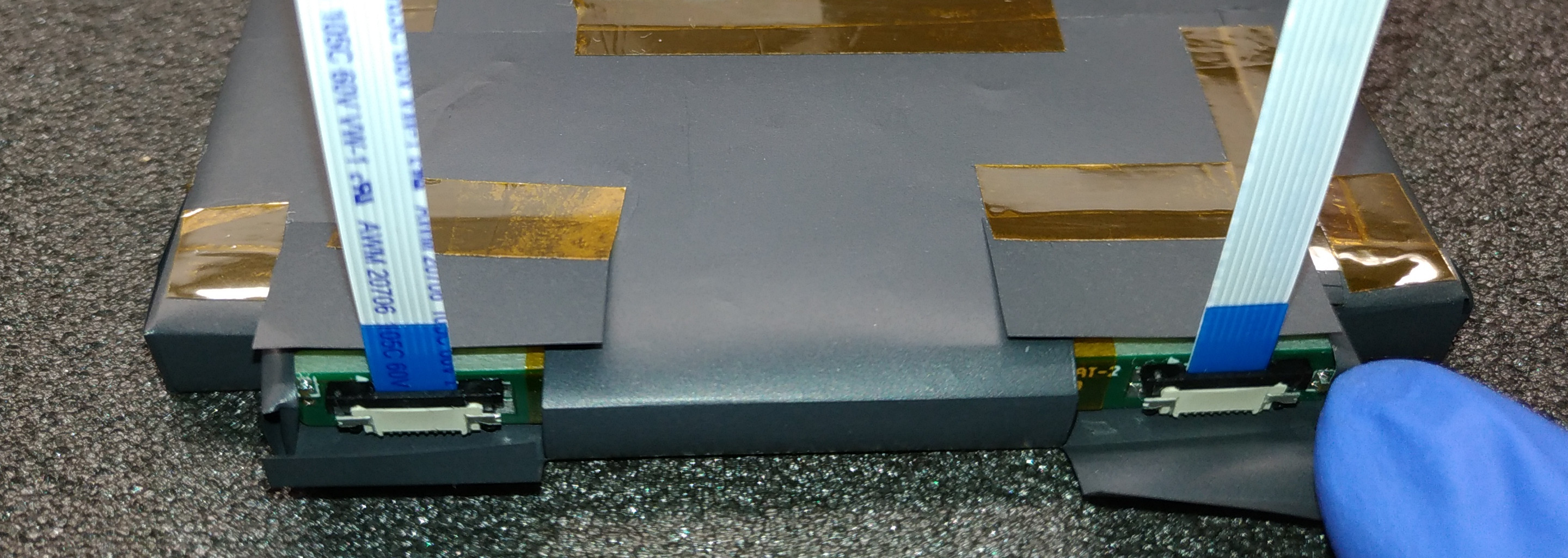}
\includegraphics[width=85mm]{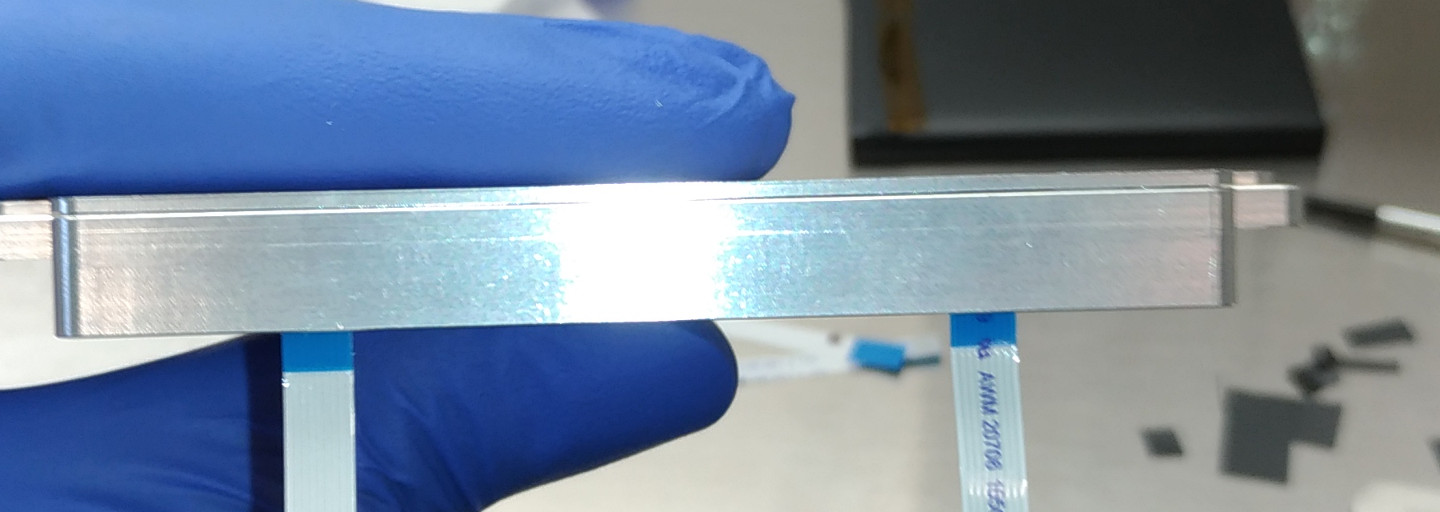}
\includegraphics[width=85mm]{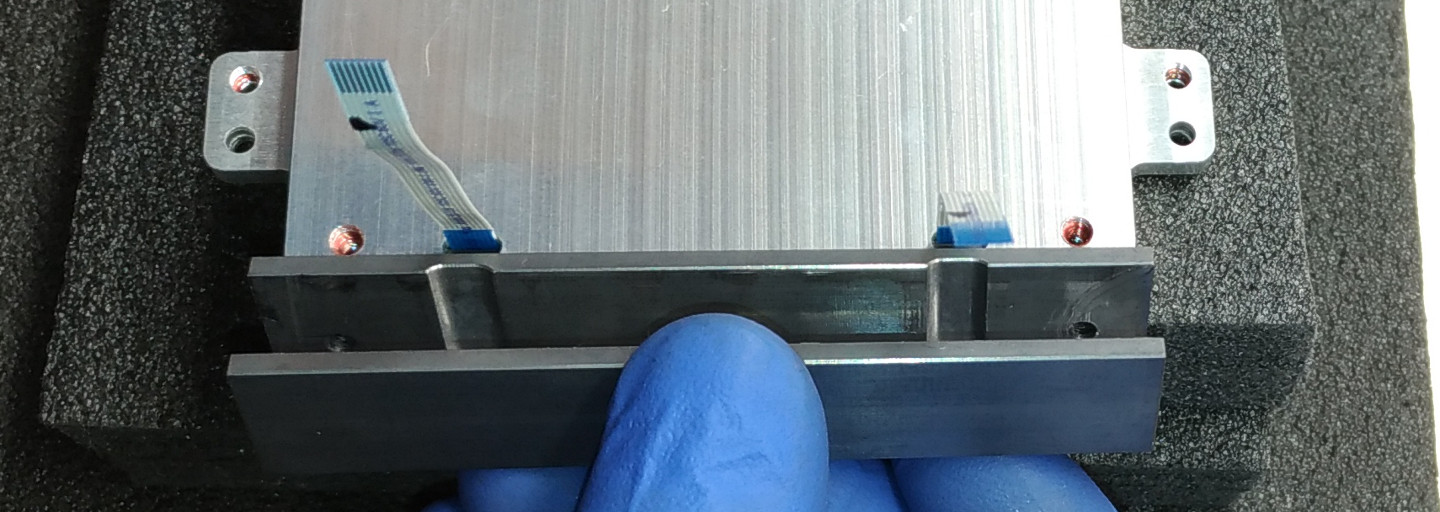}
\caption{The structure of the detector shown as a series of photos from the assembly procedure. {\it Top:} two scintillator crystals, flight model and flight spare, wrapped in ESR foil (with the exception of the positions for the MPPC arrays). {\it Second:} a PCB with two multi-pixel photon counter (MPPC by Hamamatsu) SiPM sensor arrays and thermometers. {\it Third:} the sensor array fixed onto the crystal. {\it Fourth:} one of the steps while applying the tedlar wrapping as a light trap. {\it Fifth:} the enclosure with the flex cables, side view. {\it Bottom:} mounting the lead shielding at the side of the detector where the MPPC arrays are.}
\label{fig:mppcarray}
\end{figure}

\begin{figure}
\centering
\includegraphics[width=0.95\linewidth]{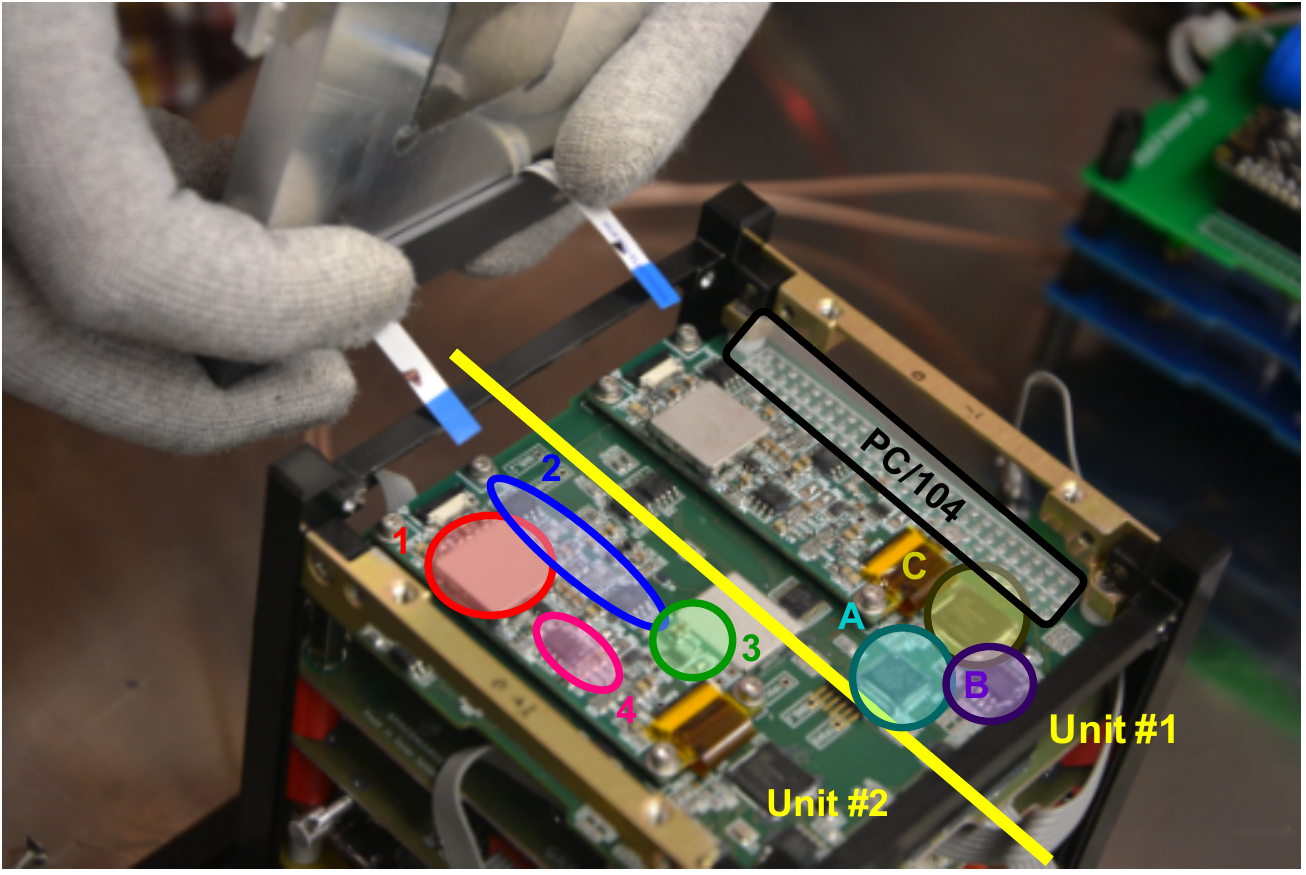}
\caption{Payload components of the \grbalpha nano-satellite. The numbered blocks mark the daughterboard containing the analog and mixed-signal components for payload unit \#2. Namely: 1 (red): adjustable high-voltage supply for MPPC reverse biasing, shielded; 2 (blue): preamplifier and signal shaping circuitry; 3 (green): analog-digital converter; 4 (magenta): high voltage control logic. The alphabetic parts mark the main PC/104 board with the digital control and signal processing parts for payload unit \#1. Namely: A (cyan): microcontroller unit; B (lilac): FPGA configuration FRAM, also used as a secondary staging area for firmware upgrades; C (yellow): FPGA that is responsible for the interfacing between the mixed-signal components on the daughterboard and the MCU. The PC/104 system bus connector can be seen at the top-right side of the figure. This photo was taken during the integration of the satellite when the scintillator block was attached electrically to the daughterboards using the white-blue flex cables. }
\label{fig:payloadcomponents}
\end{figure}

\section{Detector structure}
\label{sec:detector}

In this section we summarize the design of the satellite main payload, i.e. the scientific detector assembly and the format of the data provided by (and downloaded from) the detector electronics. The details of the integration of the detector into the satellite platform are described in Sec.~\ref{sec:satelliteplatform}.

\subsection{Scintillator and MPPCs}

As it was stated earlier in \cite{pal2020}, the core of the detector design is a thallium activated cesium-iodine crystal, having a size of $75\times75\times5\,{\rm mm}$. We applied an Enhanced Specular Reflector (ESR) wrapping around the scintillator, with the exception of a small area of the crystal where the multi-pixel photon counters (MPPCs, or MPPC silicon photomultipliers, SiPMs) are attached. A linear array of $2\times4$ S13360-3050PE MPPCs is mounted on a small, $60\times5\,{\rm mm}$ sized printed circuit board (PCB). This assembly is then wrapped into a black tedlar (DuPont TCC15BL3) layer, inhibiting stray light leaking from outside both to the detector crystal and the MPPC SiPMs. In addition to the aluminum enclosure, we mounted a lead-alloy (PbSb3) shielding at the side of the detector assembly where the MPPCs are located. The steps of the detector assembly are displayed in the panels of Fig.~\ref{fig:mppcarray}. 

\begin{figure}
\centering
\includegraphics[width=0.95\linewidth]{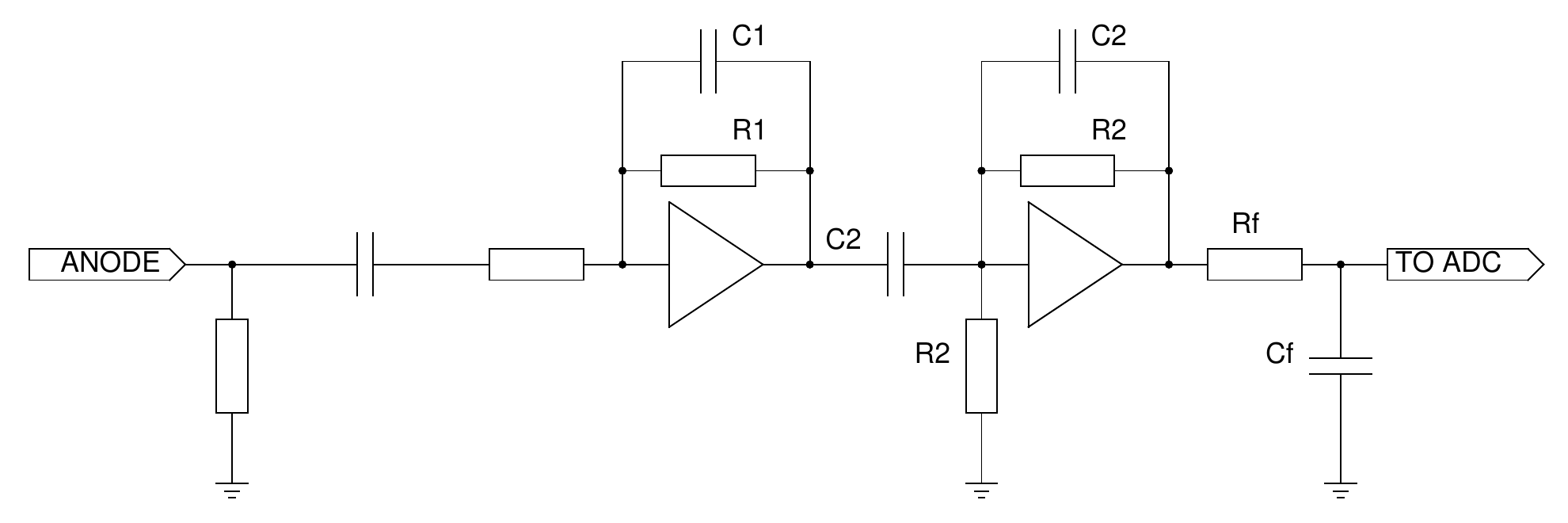}
\caption{The block-level schematics of the analog signal chain between the MPPC output and the ADC. This two-stage amplifier is formed by a traditional charge sensitive amplifier with $R_1C_1=3.3\,{\rm ms}$ decay constant and an RC-CR shaping amplifier $\tau=2.2\,{\rm\mu s}$.}
\label{fig:preamppulseshaping}
\end{figure}

\subsection{Analog and digital electronics}

The layout of detector electronics just prior to final integration is exhibited in Fig.~\ref{fig:payloadcomponents}. For having a greater flexibility in the system design, the analog frontend electronics are mounted on a separate daughterboard. On this board, the high voltage reverse bias supply is controlled via a digital-analog converter (DAC) and provides an adjustable output voltage between $45$ and $60\,{\rm V}$. Current flowing through the biased MPPCs is proportional to the amount of light detected by the photon counters. After sensed with an appropriate resistor, the signal is driven into the analog signal chain formed by a preamplifier and the pulse shaping circuitry. With the appropriately chosen resistor–capacitor (RC) networks, the pulse is widened to have a width that can fully be sampled by the analog-digital converter (ADC) without distorting its characteristics (see Fig.~\ref{fig:preamppulseshaping} for more details). Further components on the analog daughterboard are an \iic separator for the DAC and an additional power supply that provides the required voltage levels ($\pm5\,{\rm V}$) for the amplifiers and the shaping circuitry. See also \cite{torigoe2019} for further details.

Data acquisition is controlled directly by a field-programmable gate array (FPGA). The real-time processing of the FPGA is ensured by an embedded system-on-a-chip architecture where both the interface logic towards the 12-bit ADC and the communication lines towards the satellite are attached to a soft microprocessor (soft CPU). This embedded CPU allows a high level programming directly within the FPGA and it is powerful enough to run a FreeRTOS-based operating system at the same time. Communication interfaces connected to the FPGA are a full-duplex universal asynchronous receiver-transmitter (UART) and an inter-integrated circuit (\iic) bus: the primary  interface is provided by the UART line while \iic is used as a cold spare at the present implementation.

The main data acquisition mode supported by the FPGA is a dual-channel histogram accumulator. The waveform of the analog signal chain is continuously sampled with an $1.5\,{\rm\mu s}$ period (666kSPS rate) by the ADC attached to the FPGA. During the detection of a burst of optical photons, the digital logic is triggered and provides a number proportional to the energy of the incident gamma photon. The counter associated to the appropriate bin in the {\it active} histogram is then increased by one. In parallel with this accumulation, the other, {\it passive} histogram can be read out and can also be reset after data are read. At the end of the exposure, the roles of the two histogram channels are swapped within a single clock cycle, providing a truly 100\% duty cycle for the detector. In practice, the embedded block RAMs associated to both of these (otherwise identical) histograms are 32-bit wide and have a depth of 256. This setup allows a high instrumental resolution for the energy spectra as well as long exposure times without any integer overflow. On the other hand, the data transfer rate between the FPGA and the main microcontroller unit (MCU) allows exposure times as short as $20\,{\rm ms}$ even at the highest spectral resolution. 

In the current implementation, FPGA data acquisition cycles are actively controlled by a microcontroller unit. This ARM Cortex-M0 MCU core and the attached peripherals perform further processing and time-tagging of the signal, providing a temporary storage area before downlink and interfacing the detector system to the platform components such as the on-board computer and the radio transceivers. 

\begin{figure}
\centering
\includegraphics[width=0.95\linewidth]{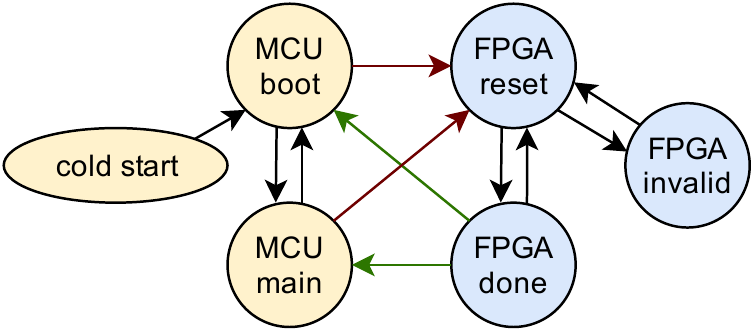}
\caption{System-level modes of operations of the \grbalpha payload: after the cold start, the microcontroller unit enters into bootloader mode but even in this mode it is capable to fully access and control the data acquisition FPGA. Once booted, regular measurements can instantly be started, however, FPGA configuration is still possible at the same time if needed. This setup allows the on-the-fly upgrade of both the MCU software and the FPGA bitstream in a safe manner while both binary images can be uploaded to the staging areas during routine operations. In the diagram, black arrows represent state transitions, red arrows denote state changes while green arrows imply state queries. Neither of the state transitions on the MCU side nor the assertion of FPGA reset state is done automatically by the system, it is only possible by telecommanding. Therefore, boot loops are not possible in this setup. If an invalid binary image is uploaded to the MCU, a watchdog reset and/or power cycling will start it again in bootloader mode, allowing the detailed examination of the situation. In the case of a failure in the FPGA bitstream upgrade, the FPGA enters to invalid mode, continuously driving its "configuration done" output low. This is detected by the MCU which can then put it back into reset state for recovery.}
\label{fig:mcufpga}
\end{figure}

Before starting routine operations (e.g. after power cycling), the payload enters to bootloader state and can only be started by sending the appropriate telecommands. This setup, along with the reconfiguration of the FPGA bitstream allows us a smooth and safe system-level operations as it is displayed in Fig.~\ref{fig:mcufpga}. Moreover, the system is capable to receive upgraded firmware images during routine operations including the cases when scientific data acquisition is ongoing. 

\subsection{Data stream}
\label{sec:datastream}

Both the on-board storage and telemetry stream employ a self-synchronizing variable-length code. Atop of the byte stream, the individual code points also form a self-synchronizing pattern at the block level, allowing decoders to unambiguously extract data even from smaller portions. This sequence is optimized for storing spectral count rates with Poisson statistics and finding weak signals above a small background. 

The symbols (code points) encoded in the raw byte stream are unsigned integers: the higher the integer number, the more number of bytes are used to store. On the other hand, this variable length encoding allows the presence of {\it overlong} sequences: within an overlong code, small numbers are stored in more bytes than the minimum number of bytes needed for storage. Table~\ref{tab:variablelengthencoding} summarizes the currently employed code space used by the payload storage system and data streams, including the aforementioned overlong sequences. For instance, the number 42 is small enough to be stored in one byte (\texttt{0x2A}) but it is allowed to be stored in two bytes (\texttt{0x40~0xAA}), three bytes (\texttt{0x60~0x80~0xAA}), etc. Such overlong characters represent block-level synchronization patterns and its code space also includes metadata about the upcoming block. As it is listed in Table~\ref{tab:variablelengthencoding}, at a maximum of 30-bit numbers are presently supported but there still is a 2-bit wide unused self-synchronization code point space for further extensions if larger integers and/or other types of data are needed to be transmitted. The rule of thumb is that one byte is needed to store 6 bits of information while data equivalent to 8 bits are interleaved for block-level synchronisation. However, those bits still encode further information regarding the type of the following block (see Table~\ref{tab:syncpatterns}, for the currently employed list of synchronization blocks).

\section{Satellite platform and system design}
\label{sec:satelliteplatform}

\begin{figure}
\centering
\includegraphics[width=0.95\linewidth]{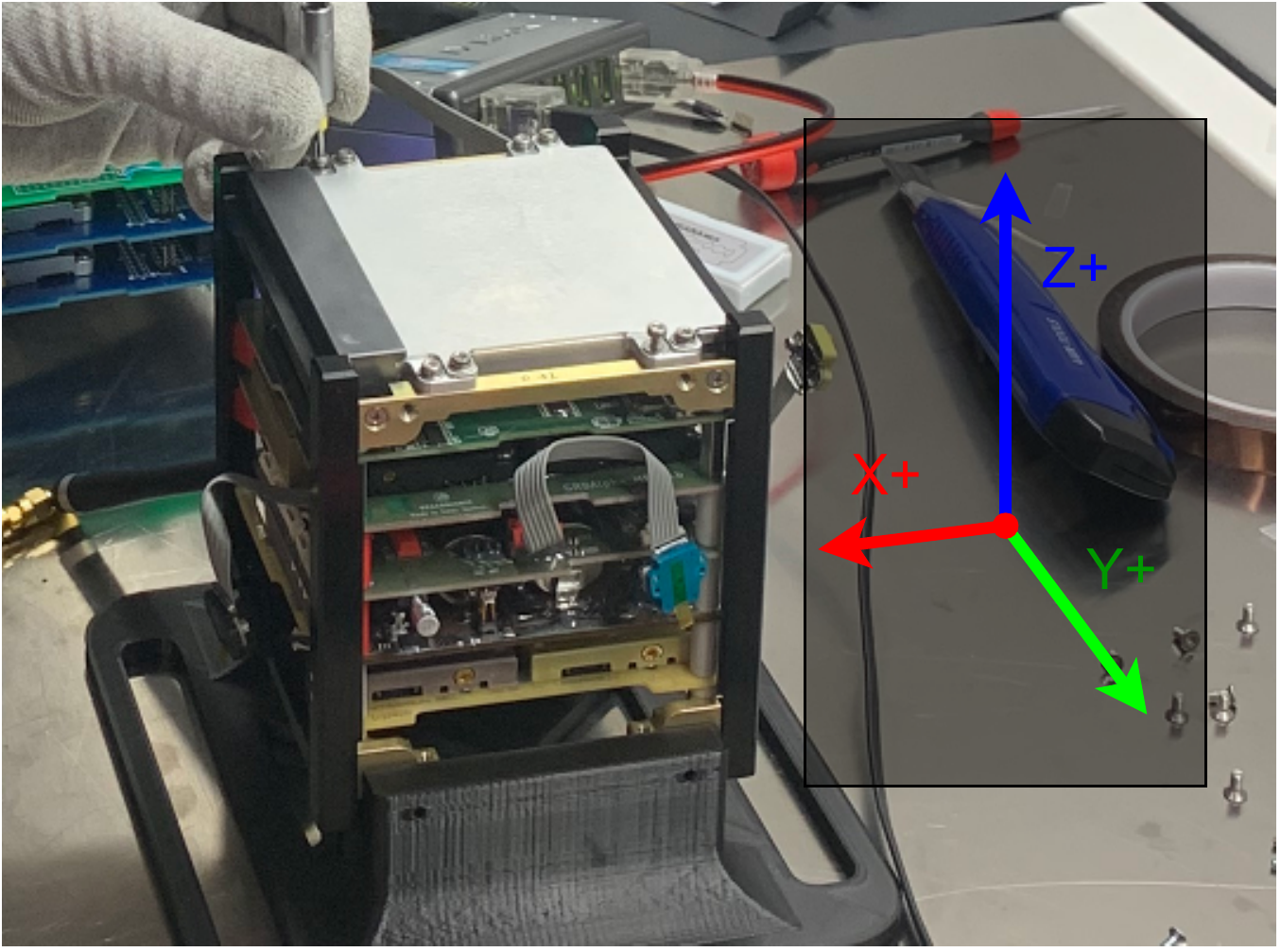}
\caption{The stack of \grbalpha and the reference frame with respect to the satellite. From top to bottom: the chasing of the gamma detector, the gamma detector payload electronics (see also Fig.~\ref{fig:payloadcomponents}), on-board computer and GNSS receiver, sensor board, power supply, radio transceivers and antenna deployer.}
\label{fig:grbalphaxyz}
\end{figure}

The mechanical construction of \grbalpha follows the standards defined by the CubeSat specifications. A 1U form factor has a dimension of $100\times100\times113.5\,{\rm mm}$, however, lateral extensions are permitted in the $X\pm$ and $Y\pm$ sides up to $6.5\,{\rm mm}$ \citep{cubesatdesign}. The full \grbalpha stack is exhibited on Fig.~\ref{fig:grbalphaxyz} along with the reference frame also involved in the detector modelling. The stack weights $1.2\,{\rm kg}$ in total and the total amount of available power averaged over one orbit is $1\,{\rm W}$. 

The primary satellite components found in the 1U-sized stack are connected using the {\it de facto} standard PC/104 connector system. This connector system distributes the power from the switchable power supply and wires the three independent internal communication interfaces between the payload electronics, on-board computer, global navigation satellite system (GNSS) receiver, sensor board, radio transceivers and the power supply. While \grbalpha is not equipped with an active attitude control system, it has permanent magnets on-board as well as patches of magnetically soft material for passive attitude stabilisation and attitude information is obtained using MEMS gyroscopes, magnetometers and sun sensors at the same time. 

Radio communication for data downlink incorporates a Gaussian frequency-shift keying (GFSK) modulation with a nominal baud rate of 9600 where individual radio packets (corresponding to one packet on the transport protocol layer) are encapsulated within the High-level Data Link Control (HDLC) framing, in accordance with the specifications defined by the AX.25 link layer. Furthermore, a linear feedback shift register is applied atop the HDLC stream with tap points defined in accordance to the G3RUH packet radio modem design in order to further whiten the GFSK radio signal and therefore allowing many $0\leftrightarrow1$ transitions for asynchronous clock recovery. Telemetry beacons are either HDLC frames or frames with additional headers defined by the AX.25 protocol. Telemetry beacons are then automatically decoded and uploaded to the public dashboard of \grbalpha\footnote{\url{https://dashboard.satnogs.org/d/iXL8Q0lGk/grbalpha}} while HDLC frames are also diverted to the console during interactive operations (see Sec.~\ref{sec:operations}). 

The uncontrolled rotation of \grbalpha provides a nearly homogeneous temperature distribution within the system. The detector temperature (see Fig.~\ref{fig:mppcarray}, second picture) varies in the range between $-5$ and $+15^{\circ}{\rm C}$ while the most exposed parts, e.g. the solar panels have a temperature within the range between $-20$ and $+25^{\circ}{\rm C}$.

\begin{figure}
\centering
\includegraphics[width=0.95\linewidth]{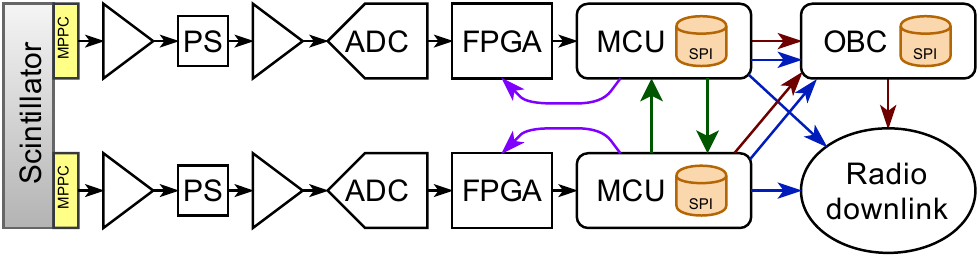}
\caption{Scientific data flow within the \grbalpha payload and platform components. Black arrows show the direction of the signal path originating from the MPPCs attached to the detector. The signal streams processed by the MCU are then being routed into various directions, depending on the currently running data acquisition configuration. Red arrows represent {\it DataKeeper} packets while blue arrows represents individual files that are retrieved either directly as file fragments or scrambled for forward error correction. File fragments can also be transferred to DK for employing the DK-based retrieval. Green arrows represent the copying functionality between the two payload nodes. This functionality allows both scientific data transfer between the two nodes (currently only for redundancy at block level) and aiding firmware upgrade by cloning either the main program binary image or the FPGA bitstream image if needed.}
\label{fig:dataflow}
\end{figure}

\begin{figure}
\centering
\includegraphics[width=0.98\linewidth]{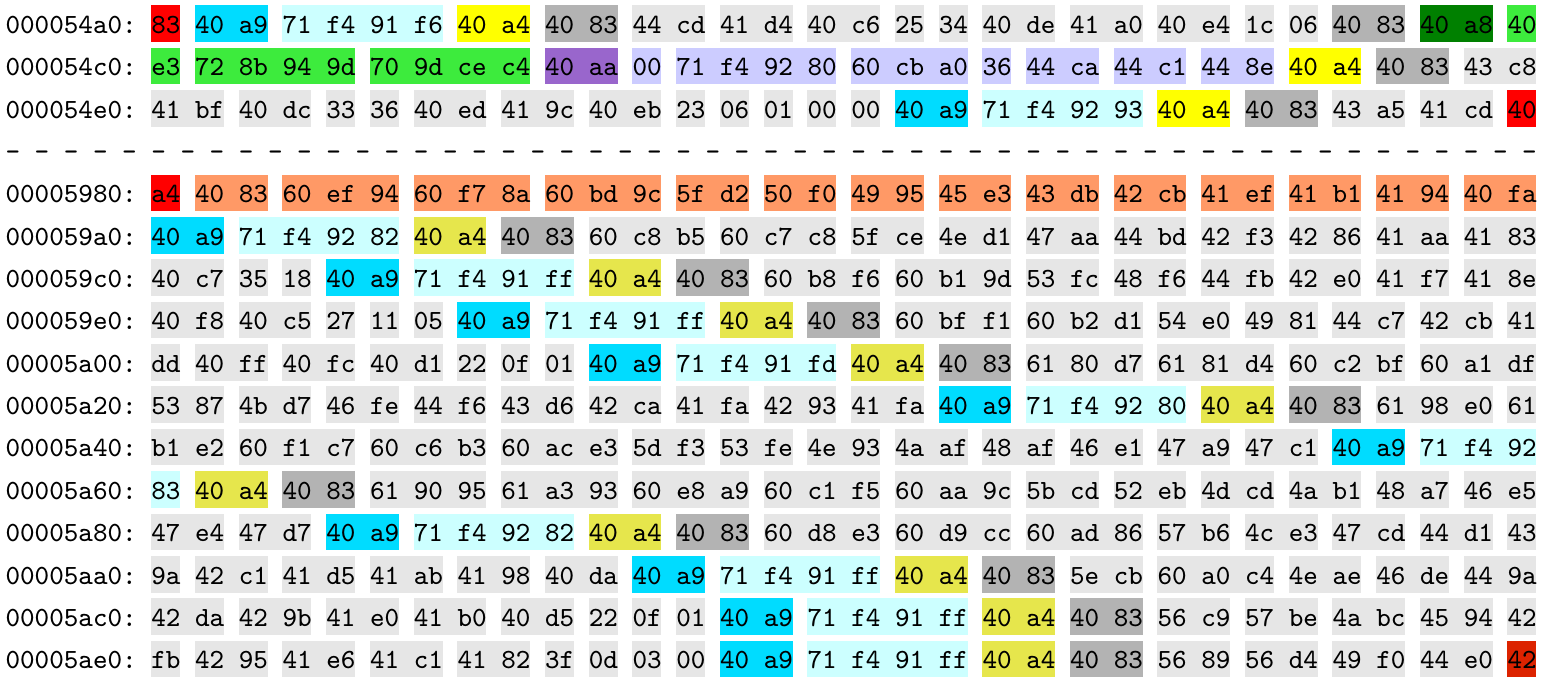}
\includegraphics[width=0.98\linewidth]{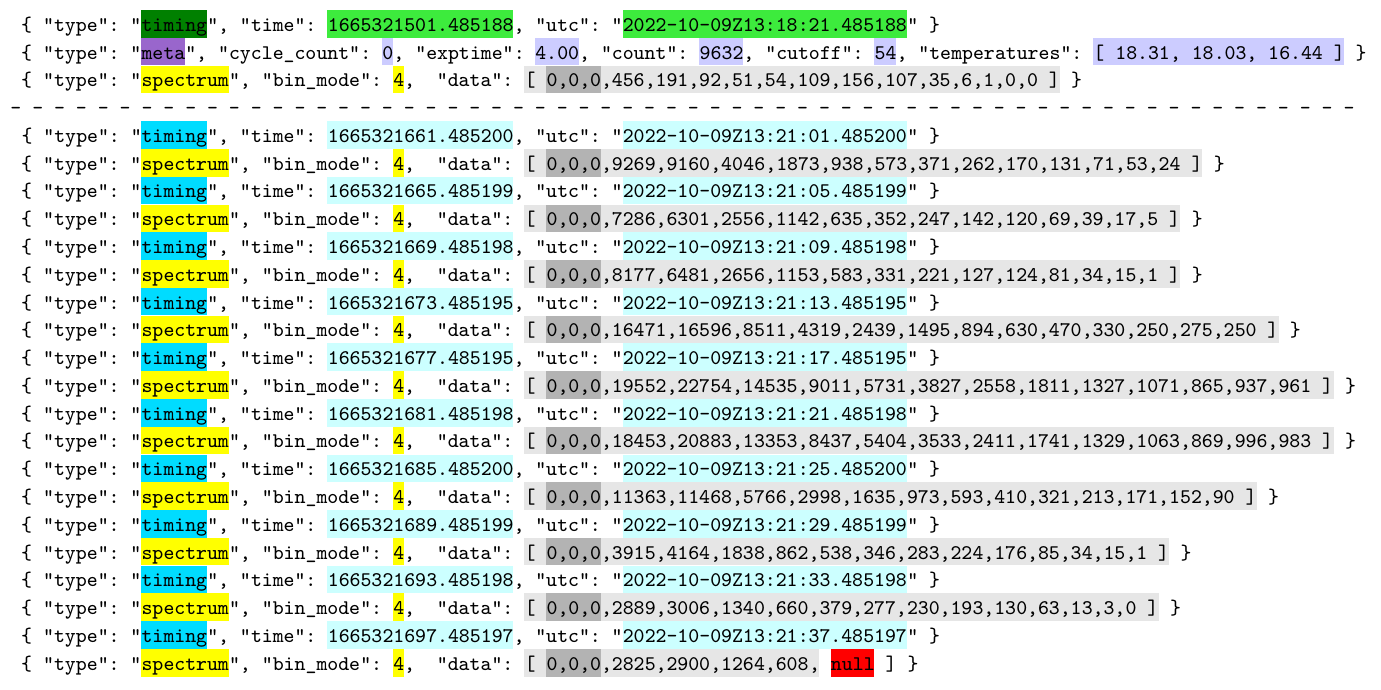}
\caption{Scientific telemetry streams from \grbalpha in raw format (upper panel) and decoded format (lower panel, displayed in the form of hierarchical JSON objects and arrays). The corresponding blocks are highlighted accordingly: absolute time instances are highlighted in green, relative time synchronization values are highlighted in blue, data acquisition metadata and housekeeping blocks are purple, spectral measurements are yellow and light gray. Bytes that are out of stream-level synchronization are highlighted with dark red while blocks are out of block-level synchronization are highlighted with orange. The trailing byte is also highlighted as red, denoting a stray byte (however, the full block can also be partially decoded, as it is clear from the JSON form). The data displayed correspond to the second peak of the GRB\,221009A event, see also the respective timestamps. Note also that relative timestamps are converted to absolute time instances during the decoding. }
\label{fig:streamexamples}
\end{figure}

\begin{figure*}
\centering
\includegraphics[width=0.45\linewidth]{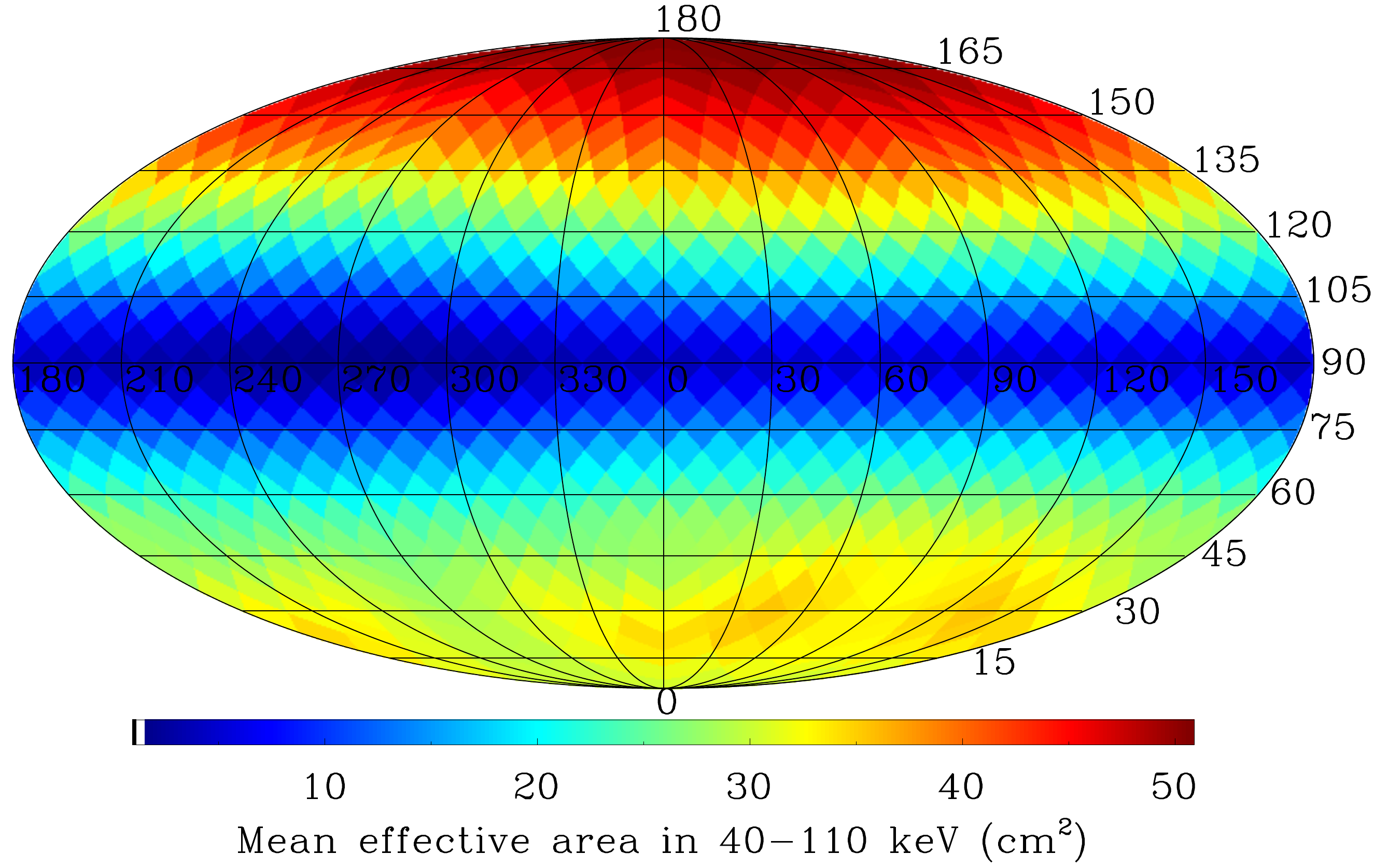}\hspace*{4mm}
\includegraphics[width=0.45\linewidth]{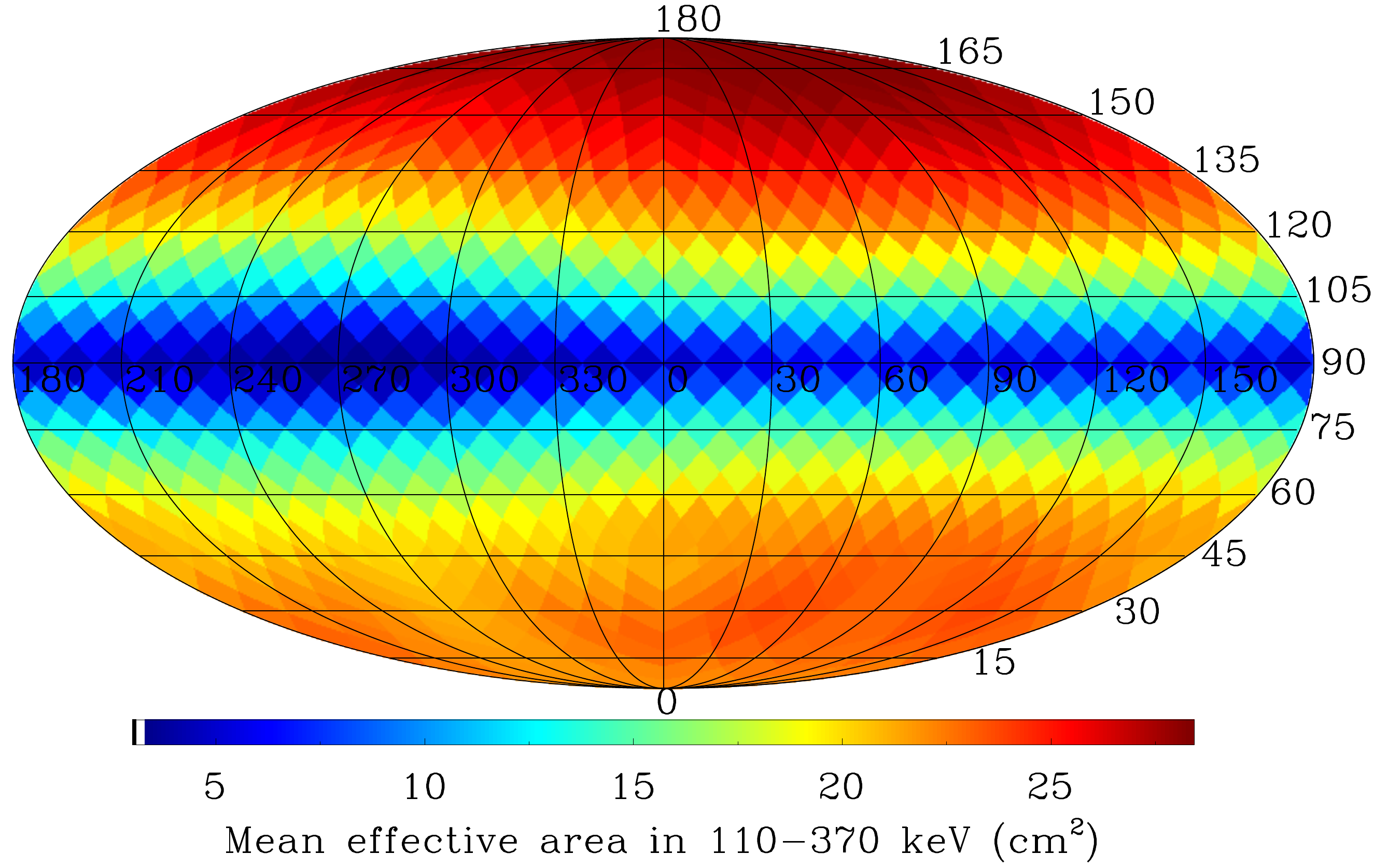}
\includegraphics[width=0.45\linewidth]{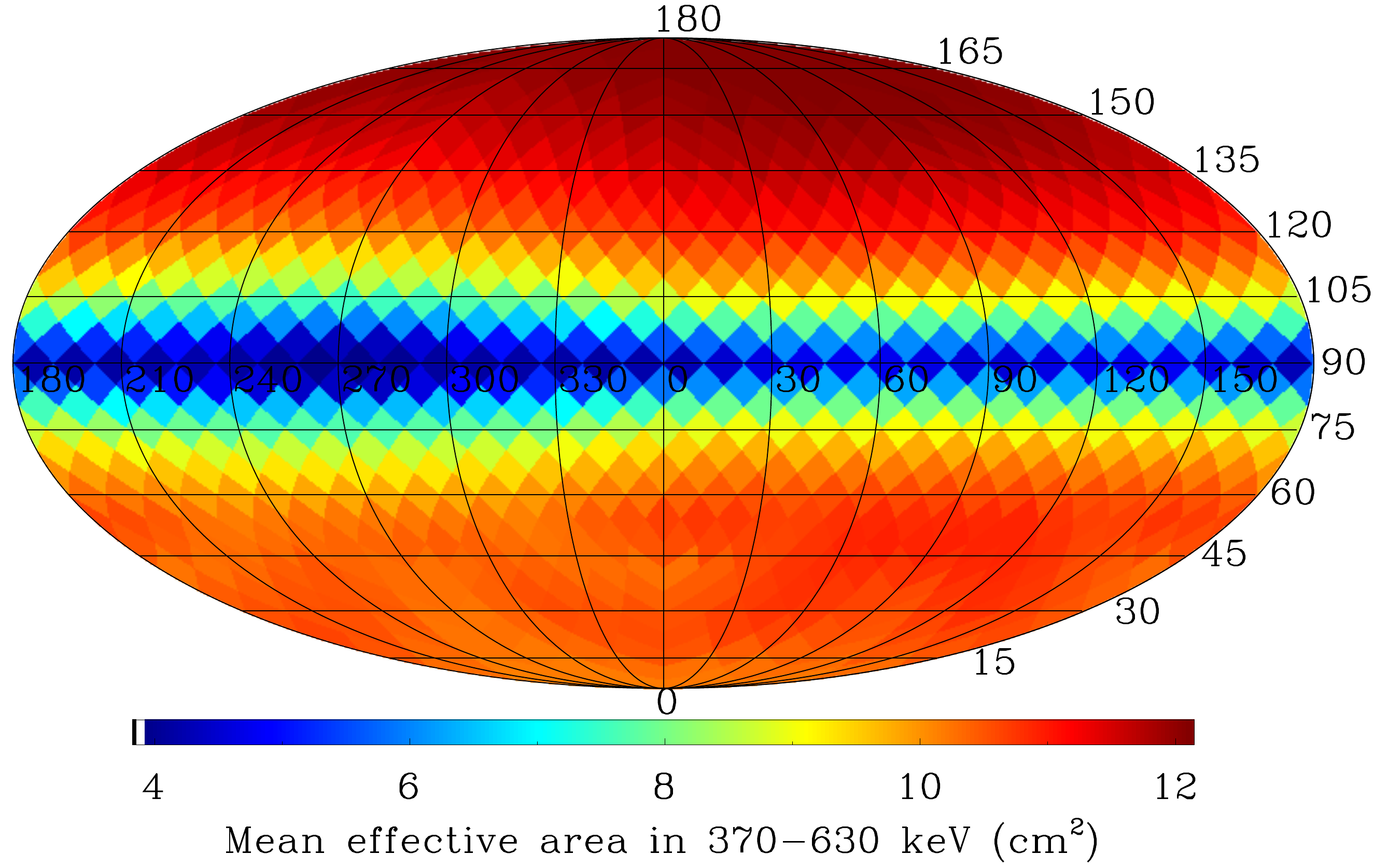}\hspace*{4mm}
\includegraphics[width=0.45\linewidth]{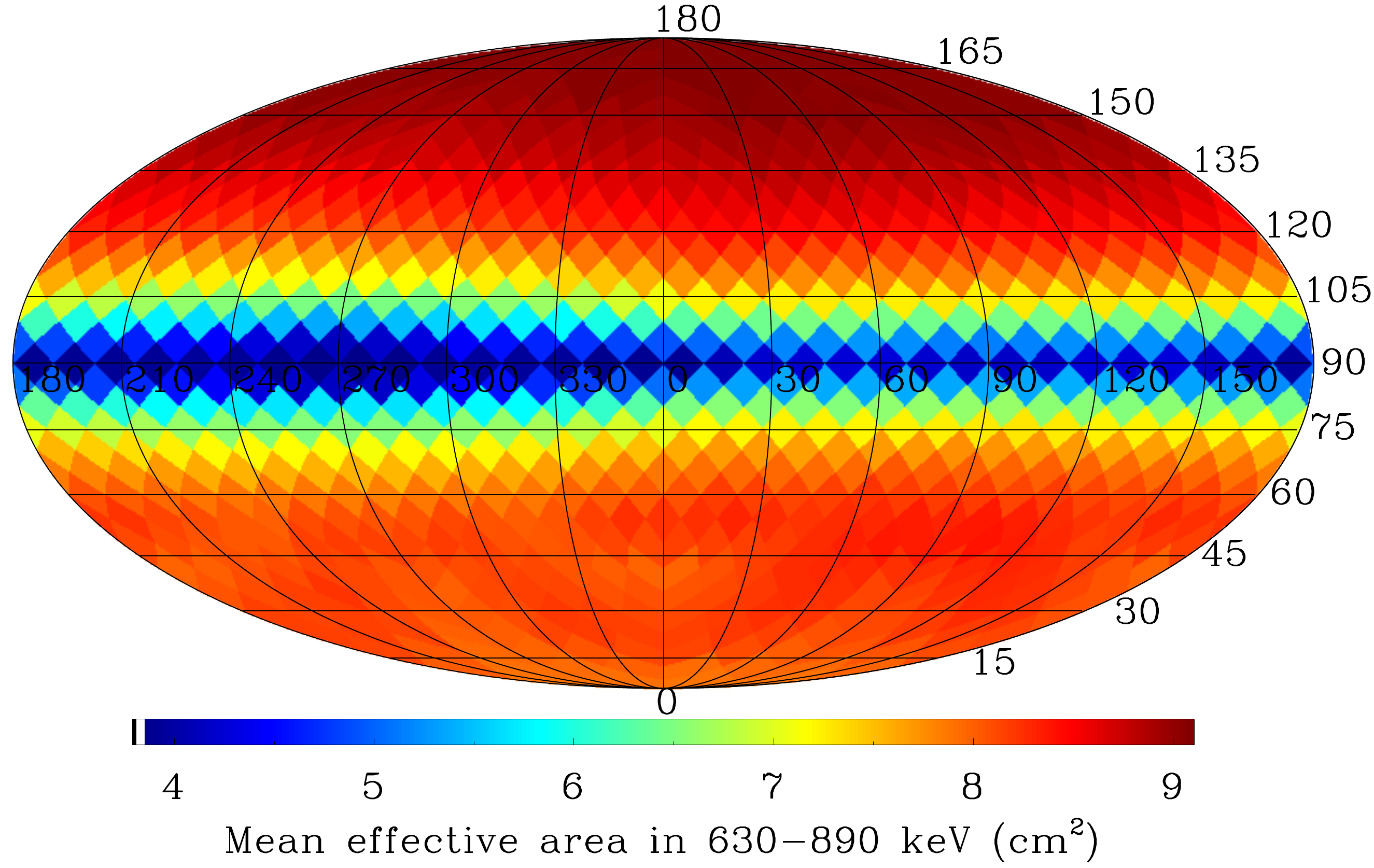}
\caption{Expected effective detector area with respect to the direction of the incident gamma radiation for various energy ranges: 
$40-110\,{\rm keV}$ (top left panel), 
$110-370\,{\rm keV}$ (top right panel), 
$370-630\,{\rm keV}$ (bottom left panel) and
$630-890\,{\rm keV}$ (bottom right panel). These $\vartheta$ (zenith angle, vertical) and $\varphi$ (azimuth angle, horizontal) plots are displayed in the reference frame of the satellite where $Z+$ axis, i.e. the face of the detector is equivalent to the $\vartheta=180^{\circ}$ angle at the top pole of spherical plots (see also Fig.~\ref{fig:grbalphaxyz}). It is clear that at lower energies, the satellite itself has a transparency around 60-70\% while at higher energies it is almost transparent and the structure of the plots is dominated by the geometric cross section of the scintillator crystal.}
\label{fig:effectiveareas}
\end{figure*}

\section{Operations and data downlink}
\label{sec:operations}

Using the currently available storage configurations, \grbalpha is operating in a semi-autonomous mode: individual observing runs are configured and queued manually during satellite contacts while data retrieval is controlled either interactively or files (and/or file fragments) scheduled for further drops above designated stations. The interactive control uses simplex stations: telecommanding is performed via an uplink station in Bankov, near Ko\v{s}ice, Slovakia while telemetry packets are received and forwarded to the console from two receiver stations found in Piszk\'estet\H{o} Observatory\footnote{\url{https://network.satnogs.org/stations/2380/}}\footnote{\url{https://ccdsh.konkoly.hu/wiki/SatNOGS_station_2380}}, Hungary and in Jablonec\footnote{\url{https://network.satnogs.org/stations/2138/}}, Slovakia. Simplex stations eliminate the need of RF power switching circuitry, greatly simplifying the station design while two receiver stations provide a nearly 100\% packet decoding during interactive sessions, compensating for the transmission fading caused by the on-board dipole antenna. During routine operations, the net scientific data downlink daily rate is around $\sim 200\,{\rm kB}$, however, with proper selection of data drops, this daily data volume could go up as high as $\sim 1\,{\rm MB}$ while still maintaining a positive power balance. 

\subsection{On-board storage}
\label{sec:storage}

\grbalpha implements two independent forms of on-board data storage schemes. First, its on-board computer allows the storage of arbitrary but small data chunks in a structure coined as {\it DataKeeper} (DK). DK is capable to store chunks received by any node on the satellite, including itself (for collecting platform-specific housekeeping data) and the payload nodes. DK has been designed to work in conjunction with packet radio based downlink and the size of the aforementioned fragments is adjusted in accordance with the maximum individual radio packet size. However, DK relies on the data link layer between the satellite and the ground station(s) during retrieval, therefore packets which are lost during the transfer are needed to be requested again if the assembly was not successful. This scheme allows simple operations, however, an excessive number of  transactions are needed to compensate the intrinsic data loss. 

In addition to DK, both nodes of the GRB payload units have their own data storage devices, allowing an independent (and optionally redundant) way of data handling. The payload firmware allocates a filesystem distributed along its storage devices, data, including routine measurements can also be stored in separate files and during downlink, files can either fully or partially be downloaded. These latter option is a preferred one during the extraction of individual GRB events (detected by other missions) where the trigger time is known. The total amount of on-board storage capacity is $2\,{\rm MB}$ for the DK while $2\times(2+64)\,{\rm MB}$ for the GRB payload units. The data storage scheme is displayed in Fig.~\ref{fig:dataflow}.

\subsection{Data downlink}
\label{sec:downlink}

Files storing scientific or auxiliary data are saved in the on-board filesystem of the payload units. Files are then transferred to ground either via the {\it DataKeeper} area of the on-board computer or directly via the radio module. Commonly, these individual files (or even portions of these files containing relevant scientific information, e.g. a few tens of minutes of recording before and after a gamma-ray burst) are too large to fit into a single AX.25 radio packet. In this case, the file $F$ is fragmented into smaller chunks, i.e. $F=\left\{f_0, f_1, \dots, f_{n-1}\right\}$ where $n$ is the total number of chunks. These chunks are, in practice, set to $k=128$ bytes, so $n=\left\lfloor(S_F+k-1)/k\right\rfloor$ where $S_F$ is the size of file $F$ in bytes. Expecting no data loss (or when these file fragments are transferred to DK), the fragments $f_i$ are transferred sequentially without any further processing.

Due to the checksum field embedded in the radio packets, a single packet is either received completely or fully discarded. Therefore, any type of packet radio forward error correction (FEC) should be implemented at a higher level. In order to add such a FEC code at packet level, the individual fragments are scrambled and converted via a partial Vandermonde transformation over the Galois field ${\rm GF}\left(2^{32}\right)$. In our practice, a file fragment $f_i$ is partitioned into a series of 32-bit unsigned integers, i.e. $f_i = \left\{ f^{(0)}_i, \dots, f^{(\ell)}_i, \dots, f^{(L-1)}_i\right\}$ where $\ell$ runs from $\ell=0$ to $L-1=31$ for a 128-byte long fragment. A packet $g_i = \left\{ g^{(0)}_i, \dots, g^{(L-1)}_i\right\}$ sent to the ground is then computed as 
\begin{equation}
g^{(\ell)}_i = \sum_{j=0}^{R-1} K_i^j\cdot f^{(\ell)}_{i+j}, \label{eq:partialvandermonde}
\end{equation}
where $R$ is the length defining the partial Vandermonde transformation and $K_i$ is a key for associated to this scrambled packet $g_i$. The multiplication involved in the computation of $K^j$ as well as during the evaluation of $K^j\cdot f^{(\ell)}_{i+j}$ is defined over the finite field ${\rm GF}\left(2^{32}\right)$ and therefore it cannot be implemented as a single binary multiply operation. If $R=1$, Eq.~(\ref{eq:partialvandermonde}) yields no additional scrambling and it is equivalent to the sequential file transfer since $K_i^j=K_i^0=1$ for all possible values of $K_i\in{\rm GF}\left(2^{32}\right)$. On the other hand, if $R=n$ and $K_i=i$, this equation is equivalent to a multiplication of the input vector with the Vandermonde matrix $V_{ij}=K_i^j=i^j$, providing full redundancy during the transfer: by receiving $n$ packets in any combination will allow the receiver to assemble the original file. However, letting $R$ as big as $n$, the computation of Eq.~(\ref{eq:partialvandermonde}) requires too much computing power: in the practice of \grbalpha operations, we use $R=n$ transfers only for files containing calibration spectra required to characterize detector degradation when the size of these files are in the order of a few kilobytes (and not hundreds of kilobytes or megabytes). During a download request, the \grbalpha payload firmware is also capable to create a randomized series of $i$ indices in order to further scramble the file transfer. Upon the reception of the $g^{(\ell)}_i$ fragments, it is both necessary and sufficient to include the length $R$, the key $K_i$, the fragment offset index $i$ and the total number of fragments $n$ (or, equivalently, the file size) within the {\it same} telemetry packet. The net size of the fragment, $L$ is simply taken from the packet size. We found this feature important due to the uncontrolled rotation of the satellite. Namely, by employing a single receiver station, the transmission could fade as long as five - ten seconds with a period of a few minutes. In this case, adjacent fragments will completely be missing from the stream even from comparatively large value of $R$ and such a scenario would make the inversion of the partial Vandermonde matrix impossible. 

For example, such a file download can be seen in the SatNOGS observation 7134188\footnote{\url{https://network.satnogs.org/observations/7134188/}}, where $\sim840$ packets have been retrieved out of the nearly $\sim 1000$ transmitted ones due to fading, however, the aforementioned scrambling and partial Vandermonde transformation with the length parameter of $R=8$ was sufficient to easily recover the $n=600$ fragments. In practice, even smaller overhead is sufficient for downlink, our experience from many hundreds of such downloads is that the minimum additional redundancy needed to be added is around $\sim15\%$, which accounts both for the packet loss due to transmission fading and the reception of fragments that are not linearly independent over ${\rm GF}\left(2^{32}\right)$. This level of redundancy is equivalent to the overhead of an RS(255,223) Reed-Solomon code. For reference, we give an implementation of the aforementioned fragment unpacking, FEC assembly and scientific data decoding on the project's website\footnote{\url{https://grbalpha.konkoly.hu/static/utils/}}. Raw packets retrieved are converted into an intermediate JSON representation (\texttt{grbalpha-downlink.sh}), which is then assembled using the FEC method described above (\texttt{asmgetf}) and then the assembled scientific streams are decoded into a standardized format (\texttt{daq-decode.sh}).

\subsection{Data products}

\refmark{By implementing the aforementioned process of retrieval described in Sec.~\ref{sec:downlink}, data are available to the community in format of a JSON representation similar to the listing displayed in Fig.~\ref{fig:streamexamples}. The count spectra in JSON files are also converted to FITS\footnote{\url{https://fits.gsfc.nasa.gov/}} files following the OGIP FITS Standards which can be used by common spectral analysis tools such as X-ray spectral fitting package XSPEC\footnote{\url{https://heasarc.gsfc.nasa.gov/xanadu/xspec/}} \citep{Arnaud1996}.
Current and typical data acquisition modes include a 1-sec cadence with 4 or 16 energy channels and calibration cycles that are run for $5\times60$ seconds with full resolution of 256 channels. Within this representation, one JSON record includes not only the photon counts for each energy bin but precise timestamps and settings related to binning are also interleaved. Another settings for the data acquisition, including exposure times, cutoff value settings (for excluding the pedestal before binning) and detector housekeeping data are interleaved with distinctive JSON object type fields. Due to the extensible nature of the JSON objects, planned data acquisition modes (e.g. parallel retrieval of data streams, for instance long exposure and high spectral resolution combined with short cadence and low resolution) can easily be inserted while still being compatible with the current data structure.}

\begin{figure}
\centering
\includegraphics[width=0.98\linewidth]{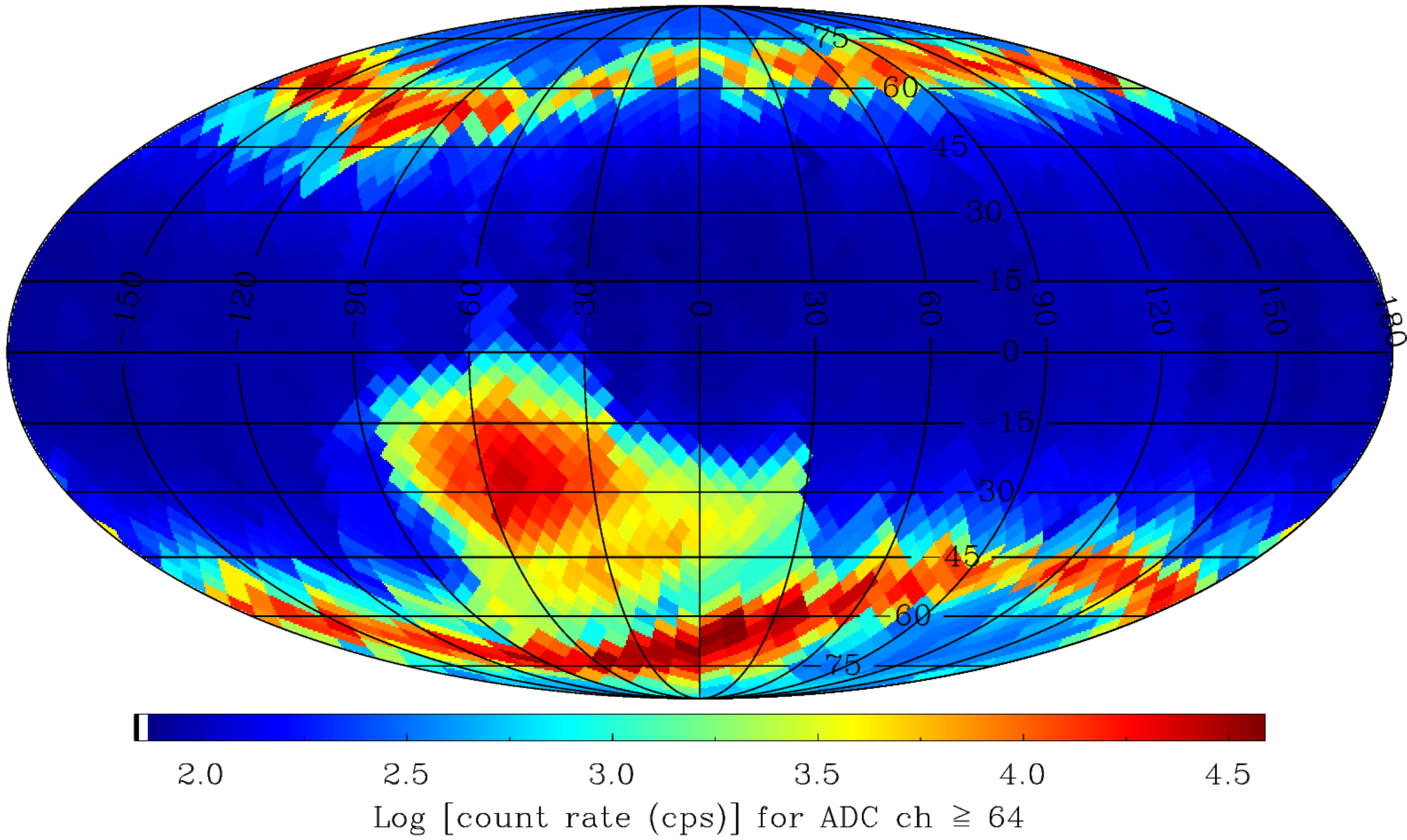}
\caption{The particle background, the extra-galactic X-ray and secondary (albedo) X-ray background radiation as measured by the detector system of \grbalpha. Data plotted here are acquired during the commissioning phase and contains nearly 2 days of continuous measurements along the orbit of \grbalpha. The northern and southern polar regions as well as the South Atlantic Anomaly are clearly visible with the elevated background levels. Otherwise, the background level is around 100 counts/sec in the full spectral range. \refmark{On this map, an equal-area Mollweide geographical projection is used where the prime meridian and the equator crosses the center.}}
\label{fig:grbbackground}
\end{figure}

\begin{figure*}
\centering
\includegraphics[width=0.98\linewidth]{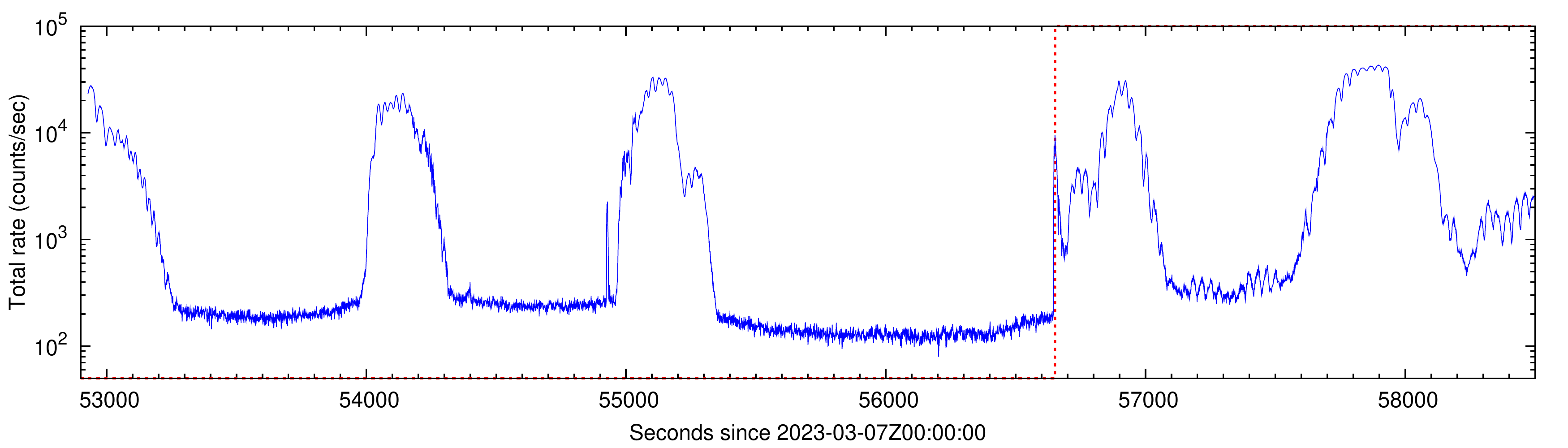}
\caption{\refmark{The light curve as observed by and retrieved from \grbalpha related to the event of GRB 230307A. The trigger has been marked with the vertical dashed red line. The total retrieved timespan was a complete orbit in this case, therefore the South Atlantic Anomaly (SAA) and four passages through polar belts are clearly visible (i.e. leaving SAA, entering and leaving the northern and southern polar rings, respectively). This prominent event has a comparable amplitude to the increased background level at the polar regions.}}
\label{fig:grb230307a}
\end{figure*}

\section{Commissioning and early scientific results}
\label{sec:commissioning}

\refmark{Along with a few dozen CubeSat-class missions and larger satellites, \grbalpha was launched on 22 March, 2021 via the support of GK Launch Services.} During commissioning, we performed all of the relevant platform-side tests of the satellite components and uploaded a revised payload firmware which was developed in between the satellite integration and the launch. Following the commissioning phase, the satellite started to perform dedicated background monitoring observations. The background level in the form of total counts per second as measured throughout the orbits of \grbalpha is displayed in Fig.~\ref{fig:grbbackground}. Based on these results, the fraction of time when the satellite is able to detect an average GRB with at lest a 5$\sigma$ significance is $\sim 67$\%. This is the duty cycle on a  550\,km polar orbit and at lower altitudes and smaller inclinations the observing efficiency is expected to be higher. \refmark{As a representative example, the light curve related to the exceptionally bright event of GRB 230307A is displayed on Fig.~\ref{fig:grb230307a}. The background-subtracted version of this light curve is shown in \cite{gcn33418}.}

\refmark{While the initial low energy threshold of the detectors were in the range of 30 keV after launch,} the degradation of the Hamamatsu MPPCs remains at an acceptable level, resulting in a low energy threshold degradation to 60-70 keV after two years of in orbit operations. A detailed evaluation of the SiPM detector degradation will be a subject of a forthcoming paper (Takahashi et al., in prep.). \refmark{In addition, we plan to perform further in-orbit adjustments of the bias voltage settings to optimize the performance at lower energies.}

If the satellite is operated continuously, the detection rate is approximately 1 transient every 5 days. The number of detected long GRBs is at the time of writing significantly higher than the number of short GRBs, which might be the result of the relatively long time bins. The initial length of time bins was 4 seconds, which was recently changed to 1 second. Further shortening of the employed bins is expected to increase the detection rate of short GRBs. 

\section{Summary and future upgrade plans}
\label{sec:summaryandplans}

Since its launch, at the time of writing, \grbalpha has detected and characterized 23 confirmed GRBs, 9 solar flares, 2 soft gamma repeaters (SGRs) and one X-ray binary outburst, including prominent events like GRB\,221009A\footnote{The list of all \grbalpha detected transients is available here \url{https://monoceros.physics.muni.cz/hea/GRBAlpha/}}. In order to further increase and extend the scientific yield, we are planning to continue to further tune and optimize the on-board software stack of the system. One of the most important short-term further upgrade plans of ours related to the payload software is the inclusion of an on-board trigger system, which would allow autonomous detection of gamma-ray transients by real time monitoring of the observed count rate. The triggering system would allow independent detection of GRBs and other events without the knowledge of the detection by other missions. Due to the availability of a GNSS receiver, we also plan to glue the GNSS output signals directly to the payload FPGA \citep{pal2018} in order to achieve a timing accuracy comparable to the on-board oscillator timing resolution. While an active attitude control is not essential for such a detector type on a small (i.e. transparent) satellite, the {\it knowledge} of the attitude is important for the proper interpretation of scientific data. \grbalpha has on-board magnetometers and sun-sensors, however, further upgrades are required in order to interleave their corresponding data within the scientific stream. Some of the free code points (see Table~\ref{tab:syncpatterns}) are reserved for this purpose. For further missions of similar needs, we developed a procedure involving thermal imaging sensors \citep{kapas2021,takatsy2022} which is also prepared for in-orbit demonstration \citep[onboard a pico-satellite platform, see][]{mrc100} and scheduled for launch in June 2023. We also plan to extend the radio telemetry beacons with scientific information in parallel with the extension of FEC within the AX.25 frame itself (and not atop AX.25, like in the case of FX.25). These extensions will be reduced to special code patterns (like Manchester code) due to the presence of bit-stuffing in the HDLC framing. 

Other nano-satellites that aim to detect GRBs and are expected to be launched in the near future include the {\it Educational Irish Research Satellite 1} (\textit{EIRSAT-1}), which will carry a gamma-ray module (GMOD), which uses SensL B-series SiPM detectors and a CeBr scintillator \citep{murphy2021,murphy2022}. A larger and more ambitious nano-satellite mission is NASA's \textit{BurstCube}, a 6U CubeSat carrying a GRB detector made of four CsI scintillators, each with an effective area of 90 cm$^{2}$ \citep{racusin2017}. Planned nano-satellite constellations include the \textit{HERMES}, which will initially consist of a fleet of 6 3U CubeSats on a low-Earth equatorial orbit. Their detector will use silicon drift detectors to detect both the X-rays from the sky and the optical photons produced in the GAGG scintillator crystals by gamma-rays \citep{fiore2020}. The Chinese student-led Gamma-Ray Integrated Detectors (\textit{GRID}) consists of GRB detectors as secondary payloads on larger 6U CubeSats. Three satellites with \textit{GRID} detectors have been launched so far and the plan is to fly the GRB detectors on further one to two dozen CubeSats \citep{wen2019}. \refmark{One of the most advantageous properties of the employment of a network of satellites is the availability for a full-sky coverage while exhibiting a functional redundancy at the same time. Although the individual satellites might have smaller effective areas compared to larger missions, the cumulative area is in the same order of magnitude. In addition, even simple geometric constraints -- is the event being obscured by Earth or not -- and attitude information -- like the proper compilation of amplitude ratios from distinct detectors being on the same or on a different spacecraft -- can help the instantaneous localization with the same hardware configuration being tested on \grbalpha now. Moreover, a timing-based localization is also feasible for such systems, exploiting proper synchronization \citep{pal2018,ohno2020,thomas2023}.}
\grbalpha itself is a precursor of the \textit{CAMELOT} constellation. We envision it to contain at least ten 3U CubeSats, each with eight times larger geometric detection area than \grbalpha \citep[see][]{werner2018,meszaros2022}.

\begin{acknowledgements}
We would like to thank our reviewer for the thorough report of our paper and all of the suggestions for improvements. Satellite components and payload development has been supported by the  KEP-7/2018 and KEP2/2020 grants of the Hungarian Academy of Sciences and the grant SA-40/2021 of the E\"otv\"os Lor\'and Research Network, Hungary. We acknowledge the support of the E\"otv\"os Lor\'and Research Network grant IF-7/2020 for providing financial background for the ground infrastructure. We are grateful to the team developing and operating the SatNOGS network as well as the operators and maintainers of the various stations, especially Brian Yeomans (2433, 2760), David Murphy and the EIRSAT-1 team (2271), Institute for Telecommunication Research, Australia (1382), Utah State University Get Away Special Team (2550) and Small Spacecraft Systems Research Center (SSSRC) at Osaka Metropolitan University (2726). We are also thankful for the many individuals and associations from the radio amateur community providing further housekeeping/telemetry packets received from our satellite. The project has also gained support by the European Union’s Horizon 2020 programme under the AHEAD2020 project (grant agreement n. 871158) and by the MUNI Award for Science and Humanities funded by the Grant Agency of Masaryk University. GD is supported by the Ghent University Special Research Funds (BOF) project BOF/STA/202009/040 and the Fonds Wetenschappelijk Onderzoek (FWO) iBOF project BOF20/IBF/124. This work was supported by the Internal Grant Agency of Brno University of Technology, project no. FEKT-S-20-6526. This research has been also supported by JSPS and HAS under Japan-Hungary Research Cooperative Program, JSPS KAKENHI Grant Number 17H06362, 19H01908, and 21KK0051.
\end{acknowledgements}

%
\bibliographystyle{aa} 
\bibliography{references.bib} 
%

\end{document}